\documentclass[prd,aps,superscriptaddress,floatfix,nofootinbib,eqsecnum]{revtex4-2}

\pdfoutput=1

\usepackage{amsmath}
\usepackage{amsfonts}
\usepackage{amsmath}
\usepackage{amssymb}
\usepackage{bm}
\usepackage{dcolumn}
\usepackage{graphicx}
\usepackage[latin1]{inputenc}
\usepackage{latexsym}
\usepackage{rotating}
\usepackage{hyperref}
\usepackage{subfigure}
\usepackage{color}
\usepackage{changes}
\usepackage{verbatim}
\usepackage{comment}
\usepackage{empheq}
\usepackage{csquotes}
\usepackage{physics}
\usepackage{float}
\usepackage{soul}
\usepackage{orcidlink}
\usepackage{amsmath,latexsym}
\usepackage{booktabs}  

\begin{document}

\title{Stark Energy Shifts due to Quantum Gravity in RGUP Algebra}

\author{Gaurav Bhandari\orcidlink{0009-0001-1797-2821}}\email{bhandarigaurav1408@gmail.com}\affiliation{Department of Physics, Lovely Professional University, Phagwara, Punjab, 144411, India}

\author{S. D. Pathak\orcidlink{0000-0001-9689-7577}}\email{shankar.23439@lpu.co.in}\affiliation{Department of Physics, Lovely Professional University, Phagwara, Punjab, 144411, India}


\begin{abstract}
In this paper, we investigate the Stark effect in the hydrogen atom under an external electric field, incorporating relativistic generalized uncertainty principle (RGUP) corrections within Minkowskian spacetime and calculate the upper bound on $\beta$ the RGUP parameter. Employing RGUP algebra and the Stetsko-Tkachuk approximation, we derive modifications to the energy spectrum for degenerate and non-degenerate states. The perturbed Hamiltonian, modified by RGUP,  enfold quantum gravitational effects. Our results reveal quantum gravitational corrections to the Stark energy spectrum in the relativistic regime, with energy shifts for non-degenerate ($n=1$) and degenerate ($n \neq 1$) cases showing additional terms proportional to $\beta$. These findings reduce to standard Stark effect results and non-relativistic GUP frameworks in the limits $\beta\rightarrow 0$ and  $c \rightarrow \infty $, establishing our model as a generalized framework for analyzing minimal length effects in relativistic quantum systems. 
\end{abstract}

\maketitle




\section{Introduction}
In recent years, quantum gravity phenomenology, particularly the study of generalized commutator brackets in quantum systems, has garnered significant attention. The generalized commutator formalism extends the Heisenberg uncertainty principle by incorporating modifications derived from various quantum gravity theories, including String/M-theory \cite{r1,r2,r3,r4,r5}, Loop Quantum Gravity (LQG) \cite{st2,st3,st4,st5,st6,st1}, and Doubly special relativity (DSR) \cite{dsr1,dsr2} which is known as Generalized Uncertainty Principle (GUP).

The GUP arises from considerations of quantum fluctuations in space-time, fundamentally limiting the simultaneous measurement of canonically conjugate variables (e.g. position and momentum etc.) with arbitrary precision. These fluctuations introduce additional uncertainties in such measurements, reinforcing the inherent limitations imposed by quantum mechanics \cite{tawfik2015review, lake2021generalised, PhysRevD.85.104029, ADLER_1999}. Quantum fluctuations at the Planck scale cause spontaneous variations in space-time geometry, rendering exact position and momentum definitions infeasible. The GUP extends the Heisenberg Uncertainty Principle by incorporating gravitational effects, imposing stricter constraints on measurement precision. The notion that gravity may influence the uncertainty principle was first proposed by Mead in 1964 \cite{mead1964possible}. As a modification of the standard Heisenberg framework, GUP has profound implications for fundamental quantum mechanics, necessitating rigorous investigations within established quantum mechanical settings to assess its viability as a quantum gravity candidate \cite{Z5,Z6,Z7,Z8}. Furthermore, several phenomenological studies have explored the effects of GUP on cosmological dynamics \cite{la,lb,G1,G2,G3}.

A key implication of GUP is the prediction of a minimal measurable length, typically of the order of the Planck length ($l_{pl} = 10^{-35}$ m), where quantum gravitational corrections become significant \cite{pl1,pl2,G7,pl3,pl4,bat}. However, incorporating a minimal length into theoretical models presents a major challenge, as it inherently breaks Lorentz invariance, given that the minimal length is not a Lorentz-invariant quantity. Consequently, most GUP formulations have been confined to non-relativistic regimes, with only limited progress in extending GUP into relativistic frameworks and Quantum Field Theory \cite{A1,cf,A2,A3,A4,A5}. More recently, efforts have been made to construct a relativistic generalization of the uncertainty principle. Notably, \cite{B1} has introduced a relativistic generalized uncertainty principle (RGUP) that preserves Lorentz invariance while reducing to the conventional GUP in the non-relativistic limit. Establishing a consistent relativistic GUP framework enables the exploration of its implications for fundamental principles and its modifications to well-established physical results.

In continuation of phenomenological investigations, we extend the analysis to the Stark effect within the broader framework of the RGUP. The Stark effect shows shift of energy level in the presence of an external electric field, which serves as an excellent candidate for investigating potential deviations induced by RGUP. High-precision spectroscopic measurements in atomic physics could offer indirect experimental evidence for generalized uncertainty principles or set constraints on quantum gravity models, making the Stark effect a promising avenue for testing fundamental theoretical frameworks.

In Sec.~\ref{back}, we provide the background mathematical framework necessary to incorporate the effects of the RGUP, enabling the recovery of minimal length corrections in the non-relativistic limit. Then, in Sec.~\ref{stark} and the subsequent subsections, we use the approach mentioned in \cite{result1,result2} to derive the RGUP-modified corrections to the Stark effect for both non-degenerate and degenerate cases, demonstrating how the formalism recovers the standard and minimal length (GUP) results. Finally, we summarize our findings in Sec.~\ref{con}.

\section{Theoretical Framework for RGUP and Minimal Length Corrections}\label{back}
The non-relativistic commutator relations incorporating the minimal observable length for $D$-dimensional space were first proposed by Kempf, Mangano, and Mann in 1995 \cite{km1,km2,km3} given by
\begin{align}
[x^i, p^j] &= i \hbar \left[ (1 + \alpha_1 p^2) \delta^{ij} + \alpha_2 p^i p^j \right], \label{d1} \\
[x^i, x^j] &= i \hbar \frac{(2\alpha_1 - \alpha_2) + (2\alpha_1 + \alpha_2) \alpha_1 p^2}{1 + \alpha_1 p^2} (p^i x^j - p^j x^i), \\
[p^i, p^j] &= 0,
\end{align}
where, $\alpha_1$ and  $\alpha_2$  are positive deformation parameters that govern the algebraic structure within the deformed space, where the indices $i, j = 0, 1, 2, \ldots, D$ . From the  Schrodinger-Robertson uncertainty relation \cite{sch1,sch2,sch3} 
\begin{equation}
(\Delta x)^2 (\Delta p)^2 \geq \frac{1}{4}\left|\langle[x,p]\rangle\right|^2,
\end{equation} 
one can ascertain the minimum uncertainty in position, expressed as
\begin{equation}
(\Delta x^i)_{\text{minimal}} = \hbar \sqrt{(D \alpha_1 + \alpha_2)}.
\end{equation}
In \cite{B2}, Stetsko and Tkachuk derived approximate expressions for the deformed position and momentum operators, neglecting higher-order corrections or, equivalently, assuming the deformation parameter to be small. To first-order accuracy in the deformation parameters, they obtained:  
\begin{align}  
x^i &= x_0^i + \frac{2\alpha_1 - \alpha_2}{4} \left( p_0^2 x_0^i + x_0^i p_0^2 \right), \\  
p^i &= p_0^i \left( 1 + \frac{\alpha_2}{2} p_0^2 \right),  
\end{align}
where the subscript ``$0$" represents the standard position and momentum operator. For the special case $\alpha_2 = 2 \alpha_1$, the position operators commute, leading to the following modified commutation relations and the modified operators as  
  
\begin{align}  
\left\{
\begin{aligned}
[x^i, p^j] &= i \hbar \left[ (1 + 2 \alpha_1 p^2) \delta^{ij} + \alpha_1 p^i p^j \right], \\  
[x^i, x^j] &= 0, \\  
[p^i, p^j] &= 0  
\end{aligned}
\right\} 
\quad \Rightarrow \quad  
\left\{
\begin{aligned}
x^i &= x_0^i, \\  
p^i &= p_0^i \left( 1 + \alpha_1 p_0^2 \right).
\end{aligned}
\right.
\end{align}

In Minkowskian spacetime with metric signature $(- + + +)$, the quadratic GUP in Eq.(\ref{d1}) takes the form as
 \begin{equation}
[x^{\mu}, p^{\nu}]=i \hbar(1+(\epsilon - \alpha)\gamma^2 p^{\rho}p_{\rho})\eta^{\mu \nu} +i \hbar (\alpha'+2\epsilon)\gamma^2 p ^{\mu}p^{\nu}, \label{1rgup}
\end{equation}
where \(\mu, \nu \in \{0,1,2,3\}\), and  \(\gamma\) is a dimensionless parameter with the inverse dimension of momentum, while \(\alpha\), \(\epsilon\), and \(\alpha'\) are also dimensionless parameters. The parameter \(\gamma\) is defined as \(\gamma = \frac{1}{M_{\text{Pl}}c}\), where \(M_{\text{Pl}}\) denotes the Planck mass. These parameters become significant in regimes where quantum-gravitational effects play a crucial role, particularly near the Planck scale. At this scale, the dimensionless parameters are generally assumed to be of order unity. For a comprehensive discussion on these parameters, see \cite{km1,C1}.

We observe that Eq.~(\ref{1rgup}) simplifies to Eq.~(\ref{d1}) in the non-relativistic limit (\(c \to \infty\)) and reduces to the standard Heisenberg uncertainty relation as \(\gamma \to 0\). Moreover, the variables \(x^{\mu}\) and \(p^{\nu}\) correspond to the physical position and momentum, which are not canonically conjugate. To address this, we introduce new four-vectors, \(x^{\mu}_0\) and \(p^{\nu}_0\), satisfying the relations  
\begin{equation}
p^{\nu}_0 = -i \hbar \frac{\partial}{\partial x_{0 \nu}}, \quad  [x^{\mu}_0, p^{\nu}_0] = i \hbar \eta ^{\mu \nu}.
\end{equation}
Following \cite{B1}, the deformed position and momentum variables, expanded up to second order in \(\gamma\), are given by  
\begin{align}
x^{\mu} &= x^{\mu}_0 - \alpha \gamma^2 p^{\rho}_0 p_{0 \rho}x^{\mu}_0 +\alpha' \gamma^2 p^{\mu}_0p^{\rho}_0 x_{0\rho}+ \xi \hbar \gamma^2p^{\mu}_0, \label{eqx}\\
p^{\mu} &= p^{\mu}_0(1+\epsilon \gamma^2p^{\rho}_0p_{0\rho}),
\end{align}
where \(\xi\) is another dimensionless parameter. The position operator satisfies the commutation relation  
\begin{equation}
[x^{\mu},x^{\nu}]= i \hbar \gamma^2 \frac{-2\alpha+\alpha'}{1+(\epsilon-\alpha)\gamma^2 p^{\rho}p_{\rho}}(x^{\mu}p^{\nu}-x^{\nu}p^{\mu}).
\end{equation}
The last two terms in Eq.~(\ref{eqx}) contain a preferred direction due to the \( p^{\mu}_0 \) term, thus introducing anisotropy and violating the principle of relativity.
To maintain relativistic invariance, we set \(\alpha' = 0\) and \(\xi = 0\). This formulation of relativistic GUP recovers the KMM algebra in the limit \(c \to \infty\). For the application to the Stark effect, we employ the approximate representation introduced by Stetsko and Tkachuk \cite{B2}, ensuring consistency with the KMM algebra in the first order by setting \(\alpha = 0\). Under the linear approximation, the modified commutation relations take the form  
\begin{align}
[x^{\mu}, p^{\nu}] &= i \hbar(1+\epsilon \gamma^2 p^{\rho}p_{\rho})\eta^{\mu \nu} + 2 i \hbar \epsilon \gamma^2 p ^{\mu}p^{\nu}, \label{a1} \\
[x^{\mu}, x^{\nu}] &= 0. \label{a2}
\end{align}
These expressions generally hold true when the GUP-corrected parameters are taken to be very small. In this regime, we can interpret \( x \) as a quasi-position operator. However, if the deformation parameters are not negligible, the neglected higher-order contributions may become significant, potentially leading to inaccuracies or deviations from the true dynamics of the system. Therefore, in situations where high precision is required or where the deformation parameters are not vanishingly small, it becomes necessary to include second-order or even higher-order terms to ensure a more accurate and reliable description.
These relations hold for the modified position and momentum operators  
\begin{align}
x^{\mu} &= x^{\mu}_0, \label{a3} \\
p^{\mu} &= p^{\mu}_0(1 + \epsilon \gamma^2 p^{\rho}_0p_{0 \rho}). \label{mp}
\end{align}
To derive the corresponding dispersion relation, we consider the squared physical momentum, which satisfies  
\begin{equation}
p^{\rho}p_{\rho} = -(mc)^2,
\end{equation}
now expressing this in terms of \(p^{\mu}_0\), we obtain  
\begin{equation}
p^{\rho}_0p_{0\rho}(1+2 \epsilon \gamma^2p^{\sigma}_0p_{0\sigma}) = -(mc)^2, \label{3dis}
\end{equation}
here, \(m\) denotes the particle's mass. Due to the presence of mixed derivatives, solving Eq.~(\ref{3dis}) analytically becomes challenging, even in spherically symmetric scenarios. Instead, we solve for \(p^{\rho}_0 p_{0\rho}\), selecting solutions that correctly reduce to \((mc)^2\) as \(\gamma \to 0\). Applying the quadratic formula, we obtain the second-order dispersion relation  
\begin{align}
p^\rho_0 p_{0\rho} &= -\frac{1}{4 \epsilon \gamma^2} + \sqrt{\frac{1}{(4 \epsilon \gamma^2)^2} - \frac{(mc)^2}{2 \epsilon \gamma^2}}, \nonumber \\
&\simeq - (mc)^2 - 2 \epsilon \gamma^2 (mc)^4 - \mathcal{O}(\gamma^4). 
\label{rmo}
\end{align}
We observe that in the non-relativistic limit (\(c \to \infty\)), the four-momentum scalar product satisfies  
\begin{equation}
p^\rho_0 p_{0\rho} = -\left(\frac{E}{c}\right)^2 + p^i_0 p_{i_0},
\end{equation}
and on taking the classical limit, we recover  
\begin{equation}
p^\rho_0 p_{0\rho} = - \hbar^2 \nabla^2. \label{b2}
\end{equation}
In the following section, we apply the RGUP framework to analyze the modified Stark effect and compute the quantum gravity-induced energy shifts in hydrogen-like atoms subjected to a uniform electrostatic field. From this point onward, we define $\epsilon \gamma^2= \beta$ and choose $\hbar=1$.

\section{RGUP-Modified Stark Effect}\label{stark}
 The Stark effect refers to the splitting of energy levels in a hydrogen-like atom when placed in an external electric field. This phenomenon arises due to the interaction between the applied field and the atomic dipole moment, leading to energy level perturbations that depend on the field strength and quantum numbers of the system. We consider a hydrogen atom with relativistic corrections applied only to the kinetic term, described by the Hamiltonian
\begin{equation}\label{H}
\mathcal{H} = \frac{p^2}{2m} - \frac{p^4}{8mc^2} + U,
\end{equation}
when subjected to a uniform external electric field along the positive \( z \)-direction, provides the interaction potential as
\begin{equation}
V = -e|\vec{E}|z.
\end{equation}
On considering the minimal length corrections in the presence of RGUP, the modification in the Hamiltonian arises from two sources, firstly, the correction in the interaction potential due to the modification of the electric field, and secondly the correction in the kinetic term due to the modification of the momentum.

We start with the Lagrangian density for an electrostatic field \cite{ar}, which is given as
\begin{align}\label{0}
\mathcal{L}&=\frac{1}{2}\epsilon_0 E(x)^2- \rho (x) \phi (x), \nonumber \\
&=\frac{1}{2}\epsilon_0(\partial^i \phi(x))(\partial^i \phi(x))- \rho (x) \phi (x).
\end{align}
where $\phi(x)$ is the electrostatic potential and $\rho(x)$ is the electrostatic charge density. 
We choose deformation only in the momentum term while the position remains fixed. The modified position and momentum operators from Eq.(\ref{mp}) in the form of derivatives are expressed as
\begin{align}\label{2}
x^\mu &\rightarrow X^\mu \equiv x^\mu_0, \nonumber \\ 
\partial^{\mu} &\rightarrow D^{\mu} \equiv \partial^{\mu}_0 \left( 1 -\beta (mc)^2 + \mathcal{O}(\gamma^4) \right).
\end{align}
Now incorporating the effects of RGUP through the modified operators Eq.(\ref{2}) into the Lagrangian density Eq.(\ref{0}), we get   \begin{align}\label{3}
\mathcal{L}_{RGUP}
&= \frac{1}{2}\epsilon_0(1-2\beta(mc)^2)E(x)^2-\rho(x)\phi(x),
 \end{align}
on comparing Eq.(\ref{3}) and Eq.(\ref{0}), we obtain the modified electric field upto the $\mathcal{O}(\beta)$ which is defined as
\begin{equation}\label{RGUPE}
    E(x)_{RGUP}= (1-\beta (mc)^2)E(x),
\end{equation}
one can easily transition to the non-relativistic regime by taking the limit \( c \to \infty \), thereby recovering the minimal length corrections as $-\beta \hbar ^2 \nabla^2 \vec{E}(x)$. The Eq.(\ref{RGUPE})  represents the modified electric field, which in turn modifies the Hamiltonian of the hydrogen atom.

On incorporating the RGUP corrections into the Hamiltonian, we noticed that we have to include both modified momentum term and the modified electric field in Eq.(\ref{H}). Therefore, the effective Hamiltonian became
\begin{equation}\label{HR}
\mathcal{H}_{RGUP}= \{1-2\beta(mc)^2\}\left\{\frac{p^2}{2m}-\frac{ p^4}{8mc^2}\right\}-e(1-\beta(mc)^2)Ez+U,
\end{equation}
this provide  the modified Stark potential after neglecting the higher order terms $\mathcal{O}(\beta^2)$ as
\begin{equation}\label{V}
V_{RGUP}= -2\beta(mc)^2\left(\frac{p^2}{2m}-\frac{p^4}{8mc^2}\right)-eEz(1-\beta(mc)^2).
\end{equation}
We use the perturbation theory to obtain the shift in the energy. On assuming the eigenstate and the unperturbed Hamiltonian are known, we can divide our analysis in two parts one of the non-degenerate case, taking only $(n=1)$, and for degenerate case $(n\neq1)$.

\subsubsection{Energy Shifts for the Ground State under RGUP Corrections ($n=1$)}
For the non-degenerate case, the energy shift is calculated using the RGUP-perturbed Hamiltonian (Eq.~\ref{HR}), where the RGUP terms are treated as perturbative corrections. The energy shift for the ground state of the hydrogen atom is then expressed as
\begin{align}\label{1e}
E_{\text{shift}} &= -2\beta (mc)^2 \left(\frac{\langle1,0,0|\nabla^2|1,0,0\rangle}{2m}  
- \frac{\langle1,0,0|\nabla^4|1,0,0\rangle}{8mc^2}\right)  - (1-\beta(mc)^2)e |E| \langle1,0,0|z|1,0,0\rangle  \notag \\
&\quad + \sum_{j\neq k} \frac{e^2|E|^2\left\langle k \right|\left[-2\beta ((mc)^2/e|E|) \left(\frac{\nabla^2}{2m} 
- \frac{\nabla^4}{8mc^2}\right) - (1-\beta(mc)^2)z\right]^2\left|j\right\rangle}{E_k^0-E_j^0}+...
\end{align}
We know that for a non-degenerate state, and following the standard results,
\begin{equation}\label{QMr}
\langle z \rangle = 0 , \quad \langle \nabla^2 \rangle = \frac{\hbar^2}{a_0^2}, \quad \langle \nabla^4 \rangle =  \frac{5 \hbar^4}{a_0^4},
\end{equation}
using the above relations, we evaluate the  linear correction in the energy shift as
\begin{equation}\label{Els}
E_{\text{shift}}= -2 \beta (mc)^2 \left(\frac{\hbar^2}{2m_e a_0^2}-\frac{5\hbar^4}{8m_ec^2 a_0^4}\right),
\end{equation}
 where $a_0$ represents the Bohr radius. One can recover the standard result that linear corrections in the energy shift is equal to zero when  $\beta \to 0$ and $c \to \infty$ in Eq.(\ref{1e}). The above linear correction does not involve \(\vec{E}\) in the energy shift expression. In the Stark effect, the modified energy shift involving \(E\) arises from the quadratic terms, assuming that cubic and higher-order terms have vanished.

The energy shift with the polarizability $\alpha_p$ of an atom is defined as 
 \begin{equation}
    \Delta_{shift} = -\frac{1}{2}\alpha_p |\Vec{E}|^2,
 \end{equation}
 for our case
 \begin{align}\label{1p}
    \alpha_p &= - 2e^2 \sum_{n \neq 1} \sum_{l,m} \frac{|\langle n,l,m|V_{RGUP}'|1,0,0 \rangle |^2}{E_1-E_n}.
 \end{align}
We have 
\begin{align}
\sum_{n \neq 1} \sum_{l,m}|\langle n,l,m|V_{RGUP}'|1,0,0 \rangle |^2 &= \sum_{n,l,m} |\langle n,l,m|V_{RGUP}'|1,0,0\rangle|^2 \nonumber \\
&= \langle 1,0,0|(V_{RGUP})^2|1,0,0 \rangle,
\end{align}
which is evaluated as
\begin{equation}
\langle 1,0,0|(V_{RGUP}')^2|1,0,0 \rangle=\langle 1,0,0|\left[-2(\beta (mc)^2/e|E|) \left(\frac{\nabla^2}{2m} 
- \frac{\nabla^4}{8mc^2}\right) - (1-\beta(mc)^2)z\right]^2|1,0,0 \rangle,
\end{equation}
 and by neglecting the higher order quantum gravitational terms $\mathcal{O}(\beta^2)$ we get
 \begin{equation}
(1-2\beta (mc)^2)\langle 1,0,0 |   z^2| 1,0,0 \rangle  + 4(\beta (mc)^2/ e E)\langle 1,0,0| z| 1,0,0 \rangle + 4 (\beta (mc)^2/e|E|)\langle 1,0,0 |  \left( \frac{\nabla ^2z}{2m}- \frac{\nabla^4 z}{8mc^2}\right)  | 1,0,0 \rangle,    
 \end{equation}
since, we know that $\langle \nabla^2 z \rangle = 
 \langle \nabla^4 z \rangle=0$ and 
 \begin{align}\label{QMr1}
    \langle z^2\rangle &= a_0^2, \\
    \langle z\rangle&=0,
 \end{align}
this gives the RGUP-modified Stark potential as
 \begin{equation}\label{2p}
|\langle V_{RGUP}'\rangle|^2=(1-2\beta (mc)^2 ) a_0^2.
 \end{equation}
We also know that the energy for nth orbit is given as
\begin{equation}
E_n= -\frac{e^2}{2a_0n^2},
\end{equation}
and we can write
\begin{equation}\label{3p}
-E_1+E_n = \frac{e^2}{2a_0}\left(1-\frac{1}{n^2}\right)\geq -E_1 +E_2 = \frac{3}{8}\frac{e^2}{a_0}.
\end{equation}
From Eqns.(\ref{1p},\ref{2p},\ref{3p}), we find the upper limit for modified polarizability parameter as
 \begin{equation}\label{ma}
\alpha_p <  (1-2\beta (mc)^2)\frac{16 a_0^3}{3},
 \end{equation}
  and the lower limit for the shift 
  \begin{equation}\label{uls}
      \Delta_{shift} > -(1-2\beta (mc)^2)\frac{8 a_0^3}{3}|\Vec{E}|^2.
  \end{equation}
It is easily observe from the above relation that we can recover the standard energy shift expression $-\frac{8a_0^3}{3}|\vec{E}|^2$ by taking the lim $\beta =0$ \cite{result1,result2}. The quantum gravitational effects introduced by the RGUP modification raise the lower limit of the energy shift.
From the modified polarizability of the hydrogen atom, we calculate the upper bound on the RGUP parameter. Using Eq. (\ref{ma}), we obtained:

\begin{equation}
\beta < \frac{1}{(mc)^2} \left( 1 - \frac{3\alpha_{p}}{16a_0^3} \right),
\end{equation}
where the experimental value of \(\alpha_p\) \cite{exp} is \(0.667 \times 10^{-30} \, \text{m}^3\), \(m = 9.1 \times 10^{-31} \, \text{kg}\), the Bohr radius is \(a_0 = 5.3 \times 10^{-11} \, \text{m}\), and \(c = 3 \times 10^8 \, \text{m/s}\). Substituting these values, we obtain the upper bound for \(\beta\) as:

\begin{equation}
\beta < 10^{42}.
\end{equation}
While the upper bound obtained here is not the tightest, it is comparable to other bounds in the literature, such as those derived from the Lamb shift \(\beta< 10^{36}\) \cite{ex1}, the scanning tunneling microscope \(\beta< 10^{21}\) \cite{ex1}, and harmonic oscillators \(\beta< 10^{6}\) \cite{ex2}. However, it is tighter than the bounds derived from Landau levels \(\beta< 10^{50}\) \cite{ex1} and other classical phenomena\cite{ex3,ex4}. It is important to note that the bounds previously derived are under the framework of GUP (Non-relativistic) corrections, while the bound here is calculated in the presence of RGUP (relativistic) corrections.

\begin{table}[h!]
\centering
\caption{Upper bounds on the GUP parameter \( \beta \) from various experiments and our RGUP result.}
\begin{tabular}{@{}ll@{}}
\toprule
\textbf{Experiment/Phenomenon} & \textbf{Upper Bound on \( \beta \)} \\ \midrule
Lamb shift                             & \( 10^{36} \) \cite{ex1} \\
Scanning Tunneling Microscope (STM)   & \( 10^{21} \) \cite{ex1} \\
Harmonic Oscillators                  & \( 10^{6} \) \cite{ex2} \\
Landau Levels                         & \( 10^{50} \) \cite{ex1} \\
Gravitational Waves (GW), Light Deflection & \( 10^{60},\ 10^{78} \) \cite{ex3,ex4} \\
\textbf{This work (RGUP, relativistic)} & \( \mathbf{10^{42}} \) \\ \bottomrule
\end{tabular}
\label{tab:GUPbounds}
\end{table}

\subsubsection{Degenerate State Energy Splitting with RGUP Modifications ($n\neq 1$)}


We know that the modified Stark potential is given by Eq. (\ref{V}). For the excited states of the hydrogen atom, specifically for the degenerate \( n = 2 \) level, there exist four degenerate states: \( |2,0,0\rangle \), \( |2,1,-1\rangle \), \( |2,1,0\rangle \), and \( |2,1,1\rangle \).  To calculate the energy shift caused by external electric field we must evaluate the matrix elements for the combination of different basis.
Since the angular momentum operator \( L_z \) commutes with \( \nabla^2 \) and \( z \), the matrix elements between states with different magnetic quantum numbers (\( m \neq m' \)) vanish. Thus, the perturbation due to the external electric field only couples specific states, leading to a block-diagonal structure in the Hamiltonian.  

The corresponding matrix elements in the basis \( \{|2,0,0\rangle, |2,1,-1\rangle, |2,1,0\rangle, |2,1,1\rangle\} \) can be computed using the expectation values of the Stark potential, incorporating the RGUP modifications. These corrections lead to a shift in energy levels, altering the standard Stark splitting in the presence of quantum gravitational effects.

The corresponding perturbed matrix is given as
\[ \langle  V_{\text{RGUP}}\rangle=
\begin{bmatrix}

\langle2,0,0|V_{\text{RGUP}}|2,0,0\rangle & 0 & \langle2,0,0|V_{\text{RGUP}}|2,1,0\rangle  & 0\\
0  & \langle2,1,-1|V_{\text{RGUP}}|2,1,-1\rangle  & 0 & 0 \\
\langle2,1,0|V_{\text{RGUP}}|2,0,0\rangle & 0 & \langle2,1,0|V_{\text{RGUP}}|2,1,0\rangle & 0 \\
0 & 0 & 0 & \langle2,1,1|V_{\text{RGUP}}|2,1,1\rangle 
\end{bmatrix}
\]

Now calculating the individual elements of the matrix using Eqns.(\ref{QMr},\ref{QMr1}) and the expectation value of $p^4$ for different $l$, we get
\begin{align}\label{def}
\langle 2,0,0 | V_{\text{RGUP}} | 2,0,0 \rangle &= -2\beta(mc)^2 \left[\frac{\hbar^2}{4a_0^2} - \frac{13}{16a_0^4} \frac{\hbar^4}{8m_ec^2} \right] \equiv -2\beta(mc)^2 M, \\
\langle 2,0,0 | V_{\text{RGUP}} | 2,1,0 \rangle &= 3a_0(1 - \beta(mc)^2)e|\vec{E}|, \\
\langle 2,1,-1 | V_{\text{RGUP}} | 2,1,-1 \rangle &= -2\beta(mc)^2 \left[\frac{\hbar^2}{4a_0^2} - \frac{7}{48a_0^4} \frac{\hbar^4}{8m_ec^2} \right] \equiv -2\beta(mc)^2 P, \\
\langle 2,1,0 | V_{\text{RGUP}} | 2,0,0 \rangle &= 3a_0(1 - \beta(mc)^2)e|\vec{E}|, \\
\langle 2,1,0 | V_{\text{RGUP}} | 2,1,0 \rangle &= -2\beta(mc)^2 \left[\frac{\hbar^2}{4a_0^2} - \frac{7}{48a_0^4} \frac{\hbar^4}{8m_ec^2} \right] \equiv -2\beta(mc)^2 P, \\
\langle 2,1,1 | V_{\text{RGUP}} | 2,1,1 \rangle &= -2\beta(mc)^2 \left[\frac{\hbar^2}{4a_0^2} - \frac{7}{48a_0^4} \frac{\hbar^4}{8m_ec^2} \right] \equiv -2\beta(mc)^2 P,
\end{align}
this gives the RGUP modified matrix as follows 
\[ \langle  V_{\text{RGUP}}\rangle=
\begin{bmatrix}

 -2 \beta (mc)^2M & 0 &  3a_0(1- \beta (mc)^2)e|\vec{E}| & 0\\
0  & -2\beta (mc)^2P  & 0 & 0 \\
 3a_0(1- \beta (mc)^2)e|\vec{E}| & 0 & -2\beta (mc)^2P & 0 \\
0 & 0 & 0 & -2\beta (mc)^2P,
\end{bmatrix}
\]
the determinant of the above matrix gives the RGUP corrections to the energy shift for degenerate case. The characteristic equation for the given matrix on neglecting the $\mathcal{O}(\beta^2)$ is obtained as  
\begin{equation}
\lambda( \lambda+4\beta (mc)^2 P  ) \left(\lambda^2+ 2 \lambda \beta (mc)^2 (M+P) -9 a^2 e^2 |\vec{E}|^2  (1 - 2 \beta (mc)^2)   \right) + \mathcal{O}(\beta^2) = 0,
\end{equation}
neglecting the higher-order corrections $\mathcal{O}(\beta^2)$, we obtain the shifts as
\begin{align}\label{ali1}
\Delta_{shift}'&=0,\\
\Delta_{shift}''&=-4\beta(mc)^2P,\\
\label{li}\Delta_{shift}'''&= \frac{-2 \beta (mc)^2 (M+P) \pm 6 e |\vec{E}|a_0\sqrt{   (1 - 2 c^2 m^2 \beta)}}{2},
\end{align}
where $M$ and $P$ are defined in Eq.(\ref{def}). We observe from Eq. (\ref{li}) that the shift contains a linear term in the electric field \( \vec{E} \), which characterizes the linear Stark effect. With the RGUP modification, additional terms involving the GUP parameter \( \beta \) appear in the shift expression. These extra terms vanish when the RGUP correction is not considered $(\beta=0)$, leaving only the standard linear dependence on \( \vec{E} \).

\section{Conclusion} \label{con}

In extreme environments of high energies where gravity becomes significant, quantum gravitational corrections become relevant. Such environments, like the intense magnetic fields surrounding neutron stars and magnetars~\cite{n1,n2,n3,a123,a1234,a12345}, are where both relativistic and quantum gravity effects play a crucial role. Investigating atomic spectra in the vicinity of these compact objects presents a valuable opportunity to explore how quantum gravity may influence spectral lines, offering a promising observational avenue to test various quantum gravity theories. In this study, we have extended the framework of the Relativistic Generalized Uncertainty Principle (RGUP) to investigate the Stark effect in a hydrogen atom, incorporating quantum gravitational corrections within Minkowskian spacetime. Our analysis demonstrates that RGUP introduces significant modifications to the energy spectrum of the hydrogen atom under an external electric field, affecting both non-degenerate and degenerate states. By employing the Stetsko and Tkachuk approximation and relativistic GUP algebra, we derived the RGUP-modified Hamiltonian and computed the resultant energy shifts.

 For the non-degenerate case ($n=1$), we obtained a linear energy shift given by:
$E_{\text{shift}} = -2 \beta (m c)^2 \left( \frac{\hbar^2}{2 m_e a_0^2} - \frac{5 \hbar^4}{8 m_e c^2 a_0^4} \right)$,
where $\beta = \epsilon \gamma^2$.  This correction is nonzero and vanishes when $\beta \rightarrow 0$, recovering the standard result where the linear shift is zero. Additionally, the quadratic Stark effect exhibits an increased lower bound for the energy shift as $\Delta_{\text{shift}} > -\left(1 - 2 \beta (m c)^2\right) \frac{8 a_0^3}{3} |\vec{E}|^2 $ i.e. RGUP raises the energy shift compared to the standard case ($\beta = 0$).

For the degenerate case ($n=2$), the RGUP-modified Stark potential yields energy shifts that include both linear and quantum gravitational terms. The characteristic equation, neglecting higher-order terms $\mathcal{O}(\beta^2)$, provides the shifts given by (\ref{ali1}), suggest additional $\beta$-dependent terms that modify the linear dependence on the electric field $\vec{E}$, vanishing when $\beta = 0$ to recover the standard linear Stark effect. We have also calculated the upper bound on the RGUP parameter of the order \(10^{42}\), which, although not the tightest found previously, is better than some quantum experiment bounds and classical phenomena. All the previous upper bounds were calculated for GUP (non-relativistic case), whereas we have calculated it for the relativistic case.

The novelty of our approach lies in developing a more generalized and comprehensive mathematical framework that captures minimal length corrections in a fully relativistic setting. This generalization is particularly important in extreme astrophysical environments, where relativistic effects are not only appropriate but necessary. The model consistently reduces to the standard Stark effect and non-relativistic GUP results in appropriate limits, affirming its generality. We also calculated the upper bound in the relativistic case, while previous bounds were obtained for non-relativistic GUP. These RGUP-induced corrections could be probed through high-precision spectroscopic measurements, offering a pathway to test quantum gravity models experimentally. This work builds on our previous RGUP analysis of the Zeeman effect and verifies the consequences of relativistic quantum gravity frameworks in understanding fundamental physical phenomena. While the RGUP modifies the Heisenberg Uncertainty Principle (HUP) at small scales and high energies, it is worth noting that a similar approach could be extended in future studies using the Extended Uncertainty Principle (EUP), which introduces modifications at large scales and incorporates a maximal length scale \cite{zhu}.



\end{document}